\documentclass[reprint,superscriptaddress,aps,prl]{revtex4-1}
\usepackage[]{notes2bib}
\usepackage{graphicx}
\usepackage{xcolor,color}
\usepackage{amsmath} % AMS Math Package
\usepackage{amsthm} % Theorem Formatting
\usepackage{amssymb}	% Math symbols such as \mathbb
\usepackage{hyperref}
\renewcommand{\v}[1]{\ensuremath{\mathbf{#1}}} % for vectors
 
\newcommand{\ang}{\mathrm{\AA}}
\let\baraccent=\= % rename builtin command \= to \baraccent

\begin{document}

\title{The Three-Dimensional Electronic Structure of LiFeAs:\\ Strong-coupling Superconductivity and Topology in the Iron Pnictides}

\author{R. P.\,Day}
\email[]{rpday7@gmail.com}
\affiliation{Department of Physics $\&$ Astronomy, University of British Columbia, Vancouver, BC V6T 1Z1, Canada}
\affiliation{Quantum Matter Institute, University of British Columbia, Vancouver, BC V6T 1Z4, Canada}
\author{M. X.\, Na}
\affiliation{Department of Physics $\&$ Astronomy, University of British Columbia, Vancouver, BC V6T 1Z1, Canada}
\affiliation{Quantum Matter Institute, University of British Columbia, Vancouver, BC V6T 1Z4, Canada}
\author{M. Zingl}
\affiliation{Center for Computational Quantum Physics, Flatiron Institute, 162 5th Avenue, New York, NY 10010, USA}
\author{B.\, Zwartsenberg}
\affiliation{Department of Physics $\&$ Astronomy, University of British Columbia, Vancouver, BC V6T 1Z1, Canada}
\affiliation{Quantum Matter Institute, University of British Columbia, Vancouver, BC V6T 1Z4, Canada}
\author{M.\, Michiardi}
\affiliation{Department of Physics $\&$ Astronomy, University of British Columbia, Vancouver, BC V6T 1Z1, Canada}
\affiliation{Quantum Matter Institute, University of British Columbia, Vancouver, BC V6T 1Z4, Canada}
\author{G.\, Levy}
\affiliation{Department of Physics $\&$ Astronomy, University of British Columbia, Vancouver, BC V6T 1Z1, Canada}
\affiliation{Quantum Matter Institute, University of British Columbia, Vancouver, BC V6T 1Z4, Canada}
\author{M.\, Schneider}
\affiliation{Department of Physics $\&$ Astronomy, University of British Columbia, Vancouver, BC V6T 1Z1, Canada}
\affiliation{Quantum Matter Institute, University of British Columbia, Vancouver, BC V6T 1Z4, Canada}
\author{D.\, Wong}
\affiliation{Department of Physics $\&$ Astronomy, University of British Columbia, Vancouver, BC V6T 1Z1, Canada}
\affiliation{Quantum Matter Institute, University of British Columbia, Vancouver, BC V6T 1Z4, Canada}
\author{P.\, Dosanjh}
\affiliation{Department of Physics $\&$ Astronomy, University of British Columbia, Vancouver, BC V6T 1Z1, Canada}
\affiliation{Quantum Matter Institute, University of British Columbia, Vancouver, BC V6T 1Z4, Canada}
\author{T. M.\, Pedersen}
\affiliation{Canadian Light Source Inc., 44 Innovation Boulevard, Saskatoon, SK S7N 2V3, Canada}
\author{S.\, Gorovikov}
\affiliation{Canadian Light Source Inc., 44 Innovation Boulevard, Saskatoon, SK S7N 2V3, Canada}
\author{S.\, Chi}
\affiliation{Department of Physics $\&$ Astronomy, University of British Columbia, Vancouver, BC V6T 1Z1, Canada}
\affiliation{Quantum Matter Institute, University of British Columbia, Vancouver, BC V6T 1Z4, Canada}
\author{R.\, Liang}
\affiliation{Department of Physics $\&$ Astronomy, University of British Columbia, Vancouver, BC V6T 1Z1, Canada}
\affiliation{Quantum Matter Institute, University of British Columbia, Vancouver, BC V6T 1Z4, Canada}
\author{W. N.\, Hardy}
\affiliation{Department of Physics $\&$ Astronomy, University of British Columbia, Vancouver, BC V6T 1Z1, Canada}
\affiliation{Quantum Matter Institute, University of British Columbia, Vancouver, BC V6T 1Z4, Canada}
\author{D. A.\, Bonn}
\affiliation{Department of Physics $\&$ Astronomy, University of British Columbia, Vancouver, BC V6T 1Z1, Canada}
\affiliation{Quantum Matter Institute, University of British Columbia, Vancouver, BC V6T 1Z4, Canada}
\author{S.\, Zhdanovich}
\affiliation{Department of Physics $\&$ Astronomy, University of British Columbia, Vancouver, BC V6T 1Z1, Canada}
\affiliation{Quantum Matter Institute, University of British Columbia, Vancouver, BC V6T 1Z4, Canada}
\author{I. S.\, Elfimov}
\affiliation{Department of Physics $\&$ Astronomy, University of British Columbia, Vancouver, BC V6T 1Z1, Canada}
\affiliation{Quantum Matter Institute, University of British Columbia, Vancouver, BC V6T 1Z4, Canada}
\author{A.\,Damascelli}
\email[]{damascelli@physics.ubc.ca}
\affiliation{Department of Physics $\&$ Astronomy, University of British Columbia, Vancouver, BC V6T 1Z1, Canada}
\affiliation{Quantum Matter Institute, University of British Columbia, Vancouver, BC V6T 1Z4, Canada}

\begin{abstract}
Amongst the iron-based superconductors, LiFeAs is unrivalled in the simplicity of its crystal structure and phase diagram. However, our understanding of this canonical compound suffers from conflict between mutually incompatible descriptions of the material's electronic structure, as derived from contradictory interpretations of the photoemission record. Here, we explore the challenge of interpretation in such experiments. By combining comprehensive photon energy- and polarization-dependent angle-resolved photoemission spectroscopy (ARPES) measurements with numerical simulations, we establish the providence of several contradictions in the present understanding of this and related materials. We identify a confluence of surface-related issues which have precluded unambiguous identification of both the number and dimensionality of the Fermi surface sheets. Ultimately, we arrive at a scenario which supports indications of topologically non-trivial states, while also being incompatible with superconductivity as a spin-fluctuation driven Fermi surface instability. 
\end{abstract}
\maketitle
\section{Introduction}
In the study of quantum materials, specific compounds within a given class are commonly identified as preferred representatives against which the majority of theories and experiments are tested. These are not necessarily those materials which optimize the desired qualities of the class, but rather those which are of unrivalled purity or simplicity. In the cuprates, this is YBCO; for topological insulators, Bi$_2$Se$_3$. In the study of iron-based superconductors (FeSC), LiFeAs plays such a role. In large part, this is owing to the crystal structure of this compound, which provides a natural, non-polar cleavage plane, allowing for preparation of atomically clean surfaces representative of the bulk crystal structure \cite{Chi_2012}. As such, this material is ideally suited to surface-sensitive techniques such as angle-resolved photoemission spectroscopy (ARPES), and scanning tunnelling microscopy (STM) \cite{Allan_2012,Chi_2012}. Together with this pristine surface, the apparent absence of magnetism, nematicity and orthorhombicity have established this stoichiometric superconductor as a focal subject and challenge in the understanding of iron-based superconductivity. However, on account of the apparent absence of Fermi surface nesting as predicted from density functional theory (DFT), LiFeAs ought to be a rather poor superconductor \cite{Mazin_2008}. Nonetheless, it supports a fairly high $T_c\sim 18$K, a fact which continues to defy explanation. Although photoemission data of exceptionally high quality and resolution has been achieved \cite{Borisenko_2012,Hajiri_2016,Brouet_2016,Fink_2019}, a survey of the reported Fermi surfaces lacks consensus, with various reports of different Fermi surface dimensionality and geometry. In connection to superconductivity, this becomes a rather important concern. In the conventional picture of iron-based superconductivity, intra-orbital antiferromagnetic spin fluctuations between hole- and electron-like Fermi surface sheets give rise to Cooper pairing with $s_{\pm}$ symmetry \cite{Mazin_2008,Graser_2009}. Models constructed on the basis of both DFT, as well as ARPES and STM, have indicated the viability of $s_{\pm}$ pairing in LiFeAs \cite{Wang_2013}. However, as we explore below, the necessary conditions identified therein are incompatible with more recent indications of a topological surface state in LiFeAs \cite{Zhang_2019}. We demonstrate that the disparity in the experimental Fermi surfaces stems largely from conflicting interpretation of spectroscopic features. More precisely, the orbital- and surface- sensitivity of the ARPES technique have obfuscated the actual low-energy electronic structure of LiFeAs. By combining polarization and photon-energy dependent experiments with bulk- and surface-projected electronic structure calculations, we identify and explain each of the contested features, presenting an unambiguous description of the Fermi surface of LiFeAs. More broadly, this work exemplifies the need to consider the vacuum interface explicitly in the interpretation of ARPES and STM studies, even on non-polar, quasi two-dimensional materials.
\begin{figure*}
	\includegraphics[width=\textwidth]{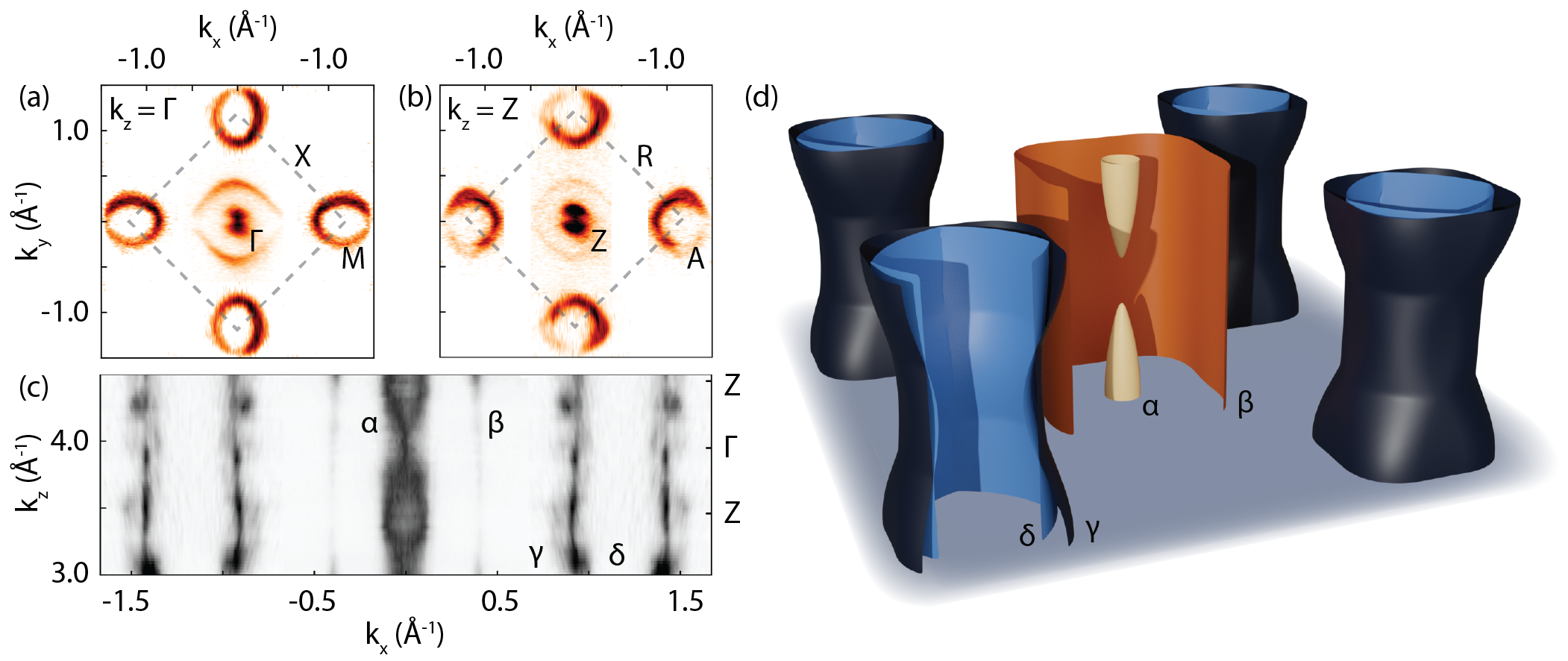}
	\caption{ARPES Fermi surface of LiFeAs: Experimental Fermi surface of LiFeAs in the (a) $k_z=0$ ($h\nu=26$ eV near $\Gamma$ and $h\nu=31$ eV near $M$), (b) $k_z=\pi/c$ ($h\nu=38$ eV near $Z$ and $h\nu=21$ eV near $A$) planes, and the (c) $k_y = 0$ plane. Panels (a,b) were acquired with linear vertical polarization (parallel to $k_y$). In (c), hole pockets $\alpha$, $\beta$ are the sum over vertical and horizontal polarizations for $h\nu \in [20, 70]$ eV; electron pockets $\gamma$ and $\delta$ were acquired with linear vertical polarization over $h\nu \in [18,76]$ eV. In panel (d), a schematic Fermi surface based on MDC fits over the 3D Brillouin zone is plotted. Electron pockets are indicated in blue, and holes in orange. The appearance of a small $\alpha$ sheet near $\Gamma$ in panel (a) is the result of SOC and residual intensity from features at lower energy, as expanded upon in Fig. 2\label{fig:figFS}}
\end{figure*}
\section{Experiment}
Single crystals of LiFeAs were grown by a self-flux method \cite{Chi_2012}, and were cleaved and measured at 25 K under ultrahigh vacuum ($<$10$^{-10}$ torr) at the Quantum Materials Spectroscopy Centre of the Canadian Light Source. Photoemission was performed over a range of photon energies from 18 - 76 eV with linear vertical and horizontal polarizations. Conversion from photon energy to $k_z$ was done using an inner potential $V_0=12.36$ eV. Samples were oriented along the Fe-Fe bond direction, allowing for $k_z$ dependent data to be acquired along the $\Gamma M$ directions at both the zone centre and corners. DFT calculations in the main text were performed using WIEN2k \cite{Blaha2001} with $RK_{max}=7$, over a shifted 20 $\times$ 20 $\times$ 20 $k$-grid and the GGA-PBE functional \cite{Perdew_1996}. We used the lattice parameters $a$ = $b$ = 3.7678 $\ang$, $c$ = 6.3151 $\ang$ from Ref. \citenum{Morozov_2010} measured at 100 K. To describe the low-energy bands crossing the Fermi energy, we constructed maximally localized Wannier functions\cite{Marzari_1997,Souza_2001} for the Fe-d shells using wien2wannier \cite{Kunes_2010} and Wannier90 \cite{Pizzi_2020}, over a 11 $\times$ 11 $\times$ 7 $k$-grid and a frozen energy window ranging from -1.87 to 3 eV. Spin-orbit coupling has been added as a local term to the Wannier Hamiltonian with coupling strength $\lambda_{SOC}$ = 40 meV. This model formed the basis of the tight-binding Hamiltonian ( detailed in the Supplementary Materials) used in the photoemission simulations. ARPES simulations were performed using the $chinook$ package\cite{Day_2019}.
\section{Results}
\begin{figure*}
\includegraphics[width=\textwidth]{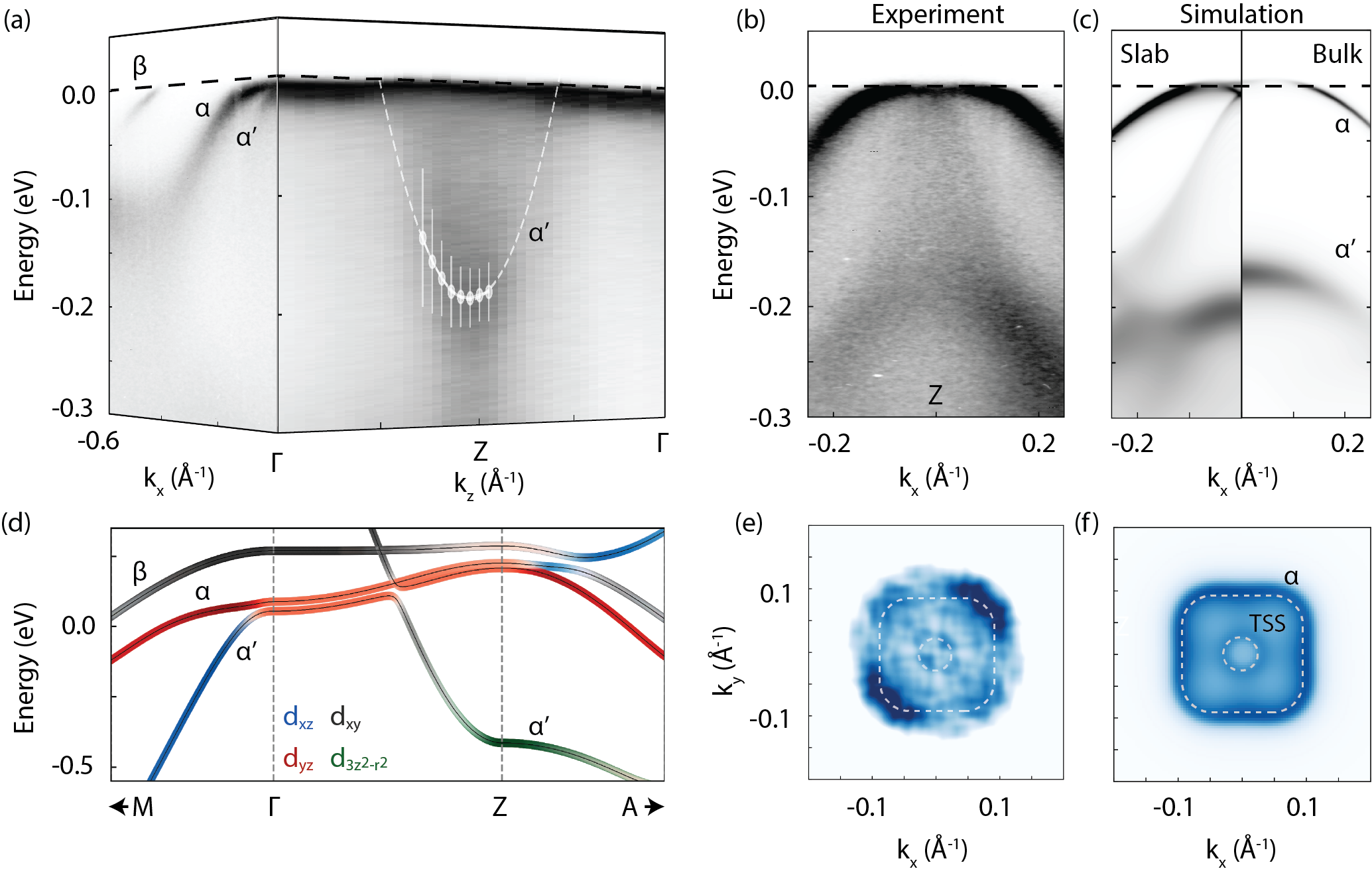}\label{fig:figGZ}
\caption{$k_z$ dispersion of the $\alpha '$ band: In (a), dispersion of hole bands along $\Gamma M$ ($h\nu=26$ eV) and $\Gamma Z$ directions  ($h\nu\in[26,54]$ eV). As anticipated from the DFT bandstructure in (d), a strongly dispersive feature near $Z$ pulls the top of $\alpha ' \sim 200$ meV below $E_F$ as indicated by EDC peak positions overlain with a parabolic fit. This seems at odds with the appearance of the smaller of two Fermi surfaces in (e); a dashed contour is superimposed as a guide. The broad hole-like feature in the spectra in (b) taken at $k_z=Z$ ($h\nu=38$ eV) corroborates this impression of a  small hole-pocket. In (c), simulated photoemission at the $Z$ point for slab (left) and bulk (right) models. The bulk simulations disregard the presence of a surface in the photoemission experiment. Comparison with experiment in (b) emphasizes the prominence of surface-related features. In particular, a linearly-dispersive topological surface state (TSS) is observed in the slab calculation near $E_F$ in panel (c). The simulated Fermi surface in (f) confirms this TSS as the progenitor of the small Fermi surface contour in (e).}
\end{figure*}

A summary of the experimental Fermi surface data, and a schematic representation based on fits to the data are presented in Fig. \ref{fig:figFS}. We identify four bulk Fermi surface sheets, corresponding to two hole pockets (ellipsoidal $\alpha$ and cylindrical $\beta$) along the $\Gamma Z$ line, and two warped cylindrical electron pockets ($\delta$ and $\gamma$) along the $M A$ line. Integrating the total volume of the Fermi surface, we confirm an occupation of 6.02(2) electrons/Fe, consistent with stoichiometric LiFeAs \bibnote{A comprehensive summary of the experimentally determined Fermi surface and band masses is provided in the Supplementary Materials Table 2}. Conspicuously absent in our summary is an innermost $\alpha'$ sheet, commonly identified as either a small 2D cylindrical, or ellipsoidal $Z$-centred hole pocket \cite{Wang_2013,Hajiri_2016,Brouet_2016}. In the absence of an $\alpha'$ pocket, this Fermi surface does not favour a weak-coupling BCS instability to superconductivity via AFM spin fluctuations. Such a situation requires a hole pocket with orbital texture matched to that of the electron pocket at the zone corner, a texture that only the $\alpha'$ sheet could provide; as the orbital texture of the $\alpha$ pocket is phase-rotated by $\pi/2$ about the $k_z$ axis, it cannot explain the incommensurate peaks observed via inelastic neutron scattering \cite{Qureshi_2014,Li_2016}. A summary of the band structure with dominant orbital contributions, as predicted from DFT, is provided in Fig. \ref{fig:figGZ}(d) for reference.

\begin{figure}
\begin{center}
\includegraphics[width=\columnwidth]{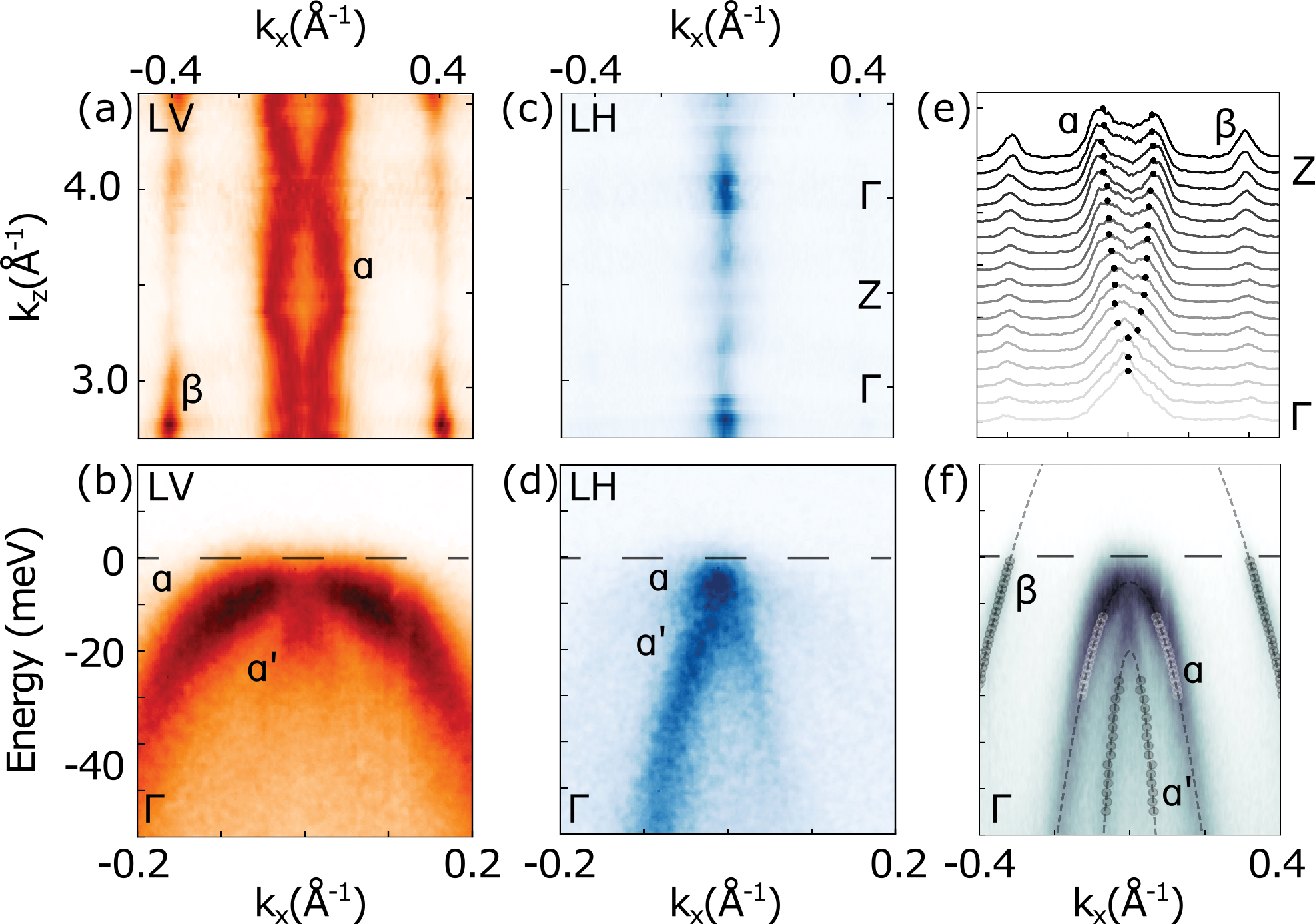}
	\caption{Spin-orbit coupling effects on apparent Fermi surface of LiFeAs. In (a,b), constant energy contours at $E=0$ eV in the $k_x$-$k_z$ plane, acquired with linear vertical (a), and horizontal (b) polarizations over $h\nu\in[20,70]$ eV. In (c,d), a single spectrum at $k_z=0$ ($h\nu=26$ eV) under the same experimental configurations as in (a,b), respectively. Although LV data suggest a warped cylindrical $\alpha$ Fermi surface sheet, a closed 3D Fermi surface is evident from composite data including both channels. In (e), MDCs at $E=0$ for the sum over linear vertical, horizontal polarizations, along the $k_z$ axis. MDC peak fits are indicated by red cursors. In (f), a composite image of both polarizations, along with MDC fits, indicating the $\alpha$ band maximum at $E=-5$ meV at $k_z=0$. \label{fig:figSOC}}
\end{center}
\end{figure}

In the ARPES literature, the $\alpha'$ sheet is reported to be nearly concentric within the larger $\alpha$ pocket. Whether both states support a Fermi surface, and whether such surfaces are open or closed along $k_z$, is the subject of controversy. As we will show, much of the ambiguity is due to the compounded effects of spin-orbit coupling and the polarization dependence of ARPES spectra. The photoemission matrix element endows the technique with a certain orbital sensitivity, a quality which has been leveraged to identify orbital texture in the FeSCs \cite{Wang_2012,Hajiri_2016,Watson_2017,Day_2019}. In the present case, this quality is a liability, and has led to the erroneous identification of the $\alpha$ sheet as a corrugated 2D cylinder, inconsistent with the 3D $Z$-centred pocket in Fig. \ref{fig:figFS}(c,d) and identified in quantum oscillations \cite{Zeng_2013}. Photoemission selection rules require acquisition of complementary datasets with orthogonal polarizations to confirm the absence of the $\alpha$-sheet at $\Gamma$. As illustrated by Fig. \ref{fig:figSOC}(a-d), spin-orbit coupling \cite{Borisenko_2016,Vafek_2017,Day_2018} mixes orbital character near $k_{||}$ = 0, with bands of $d_{xz/yz}$ character evolving to $d_{xz}\pm id_{yz}$ near $k_{||}=0$. In the photoemission data, the result is an apparent depletion of spectral weight at the $\alpha$ band maximum, for linear-vertical polarized light. Measuring with the orthogonal linear-horizontal polarization, one recovers this lost spectral weight. By combining spectra with both polarizations (analogous to measurements with unpolarized light), as in Fig. \ref{fig:figSOC}(f), one observes a continuous evolution of the band over the full range of momentum. From this composite data, we confirm that the $\alpha$ band has a maximum at $E = -5$ meV at $k_z$ = 0. However, this band does form a small $Z$-centred pocket which extends out to $k_F^z$ = 0.40(2) $\ang^{-1}$, as indicated by the merging of MDC peaks in Fig. \ref{fig:figSOC}(e). 

As the $\alpha$ band is then below $E_F$ at $\Gamma$, so too is the $\alpha'$ sheet. This much is clear from the spectra in Fig. \ref{fig:figSOC}(f), where a parabolic fit to MDC peaks indicates a band maximum at $E$ = -20(1) meV. Proceeding along $k_z$ from $\Gamma$ to $Z$, the fate of the $\alpha'$ band becomes somewhat ambiguous. From DFT, one expects an SOC-derived hybridization with a sharply-dispersing out-of-plane state, as illustrated in Fig. \ref{fig:figGZ}(d), where we have projected the DFT solution onto Wannier functions to produce a Fe 3d tight-binding model \cite{Mostofi_2014}. We note that the avoided crossing is observed for both 10 orbital (Fe 3d only), and 16 orbital (Fe 3d + As 4p) models. The pairs of states can be labelled as odd and even parity even in the 10 orbital projection implemented here, as the dispersive band is an odd-parity linear combination of the two-inequivalent Fe $3d_{3z^2-r^2}$ orbitals in the two iron unit cell. The parity eigenvalues of the associated bands at the time-reversal invariant points support the evolution of a topological surface state in either choice of orbital subspace\cite{Fu_2008}. In contrast to the situation at $\Gamma$, moving along the $k_z$ axis to $Z$, the SOC-driven hybridization and band inversion leads to an electron-band of $d_{xz/yz}$ character above $E_F$, and a $d_{3z^2-r^2}$ hole band well below $E_F$. This becomes apparent when comparing the $M\Gamma$ and $ZA$ panels of Fig. \ref{fig:figGZ}(d). Although this implies there is no $\alpha'$ Fermi surface sheet, such a $k_z$ dispersion has been identified as the source of a possible topological surface state (TSS) \cite{Wang_2015,Zhang_2019,Songtian_2020}. This scenario is supported by data acquired along $k_z$ at $k_{||} = 0$, plot in Fig. \ref{fig:figGZ}(a). Fits to the EDCs reveal the dispersive band along $k_z$. However, individual spectra at $Z$, as in Fig. \ref{fig:figGZ}(b) suggest the persistence of a hole band, up to $E_F$. Moreover, the Fermi surface in Fig. \ref{fig:figGZ}(e) corroborates previous experiments which have identified a $Z$-centred $\alpha'$ sheet \cite{Wang_2013,Hajiri_2016,Brouet_2016}. Models based on these experiments have resolved this contradiction by pushing the out-of-plane orbitals far from $E_F$ with a large on-site term \cite{Wang_2013,Saito_2014,Autore_2014,Ahn_2014}. Doing so, one can construct a model compatible with weak-coupling $s_{\pm}$ pairing: intra-orbital scattering at $q\sim(\pi/a,\pi/a)$ is recovered with a small $Z$-centred $\alpha'$ pocket. However, with the 3$d_{3z^2-r^2}$ state now above $E_F$, there is no possibility for a band inversion with the even-parity-$\alpha'$ band below $E_F$. This would preclude an explanation for recent reports of a topological surface state in LiFe$_{1-x}$Co$_x$As \cite{Zhang_2019}. How can one reconcile the apparent dispersive feature in Fig. \ref{fig:figGZ}(a) with the band in Fig. \ref{fig:figGZ}(b) and its associated Fermi surface in Fig. \ref{fig:figGZ}(e)? A satisfactory explanation of these spectral features can be had through consideration of the surface-sensitivity of ARPES; the complicated and apparently conflicting features observed in Fig. \ref{fig:figGZ} can be identified as the confluence of several surface-related issues.

The mean free path ($\lambda$) of photoelectrons is expected to be of the order of $5-10\ang$ in the ultraviolet regime, implying an acute surface sensitivity in photoemission experiments \cite{Seah_1979}. Heuristic arguments approximate surface-sensitivity in ARPES as equivalent to a Lorentzian broadening over the $k_z$ axis of the bulk electronic structure, with a width of $1/\lambda$ \cite{Hedin_2001,Strocov_2003}.  The uncertainty arguments used to arrive at this form disregard the nature of the vacuum interface which gives rise to $k_z$ uncertainty, ultimately failing to capture the functional form and magnitude of the $k_z$ broadening, as well as shifts in the spectral peaks \cite{Day_2021}. To better represent actual experiments, one ought to consider the surface explicitly \cite{Zhu_2013}. Beginning with a bulk model Hamiltonian, one can construct a large slab unit cell, suitably converged to the limit of a semi-infinite crystal and terminated with a vacuum buffer \cite{Sun_2013}. One can then model the photoemission intensity following the conventional strategies \cite{Zhu_2013,Moser_2017,Day_2019}. Surface sensitivity is implemented through exponential attenuation of contributions from sub-surface orbital basis states. As detailed in Ref. \citenum{Day_2021}, escape-depth ($\lambda$) dependent $k_z$-broadening emerges naturally in this framework, converging to the Lorentzian description in the large $\lambda$ limit. At the same time, one recovers contributions from surface states, and surface-modified bulk states. Moreover, this realistic context enables study of peak-shifts and apparent bandwidth renormalizations which arise from surface-sensitivity alone. Comparing the bulk- and surface- projected models of LiFeAs ARPES spectra at $Z$ in Fig. \ref{fig:figGZ}(c), we identify spectral features in Fig. \ref{fig:figGZ}(b) as having originated from the surface-sensitivity of the photoemission process. Specifically, the intensity near $k_{||}=0$ is the combination of $k_z$-integration over the dispersive $d_{xz/yz}/d_{3z^2-r^2}$ state, and the topological surface state reported recently \cite{Zhang_2019}. We further confirm the small Fermi surface at $Z$ seen in Fig. \ref{fig:figGZ}(e) as that of the topological surface state with our calculated surface-projected ARPES Fermi surface in Fig. \ref{fig:figGZ}(f). Large and sharp intensity within approximately 10 meV of $E_F$ is attributed to the TSS, and the remaining broad intensity which extends several hundred meV is derived from $k_z$ integration over dispersive bulk states. These measurements and calculations then confirm the absence of an $\alpha'$ Fermi surface in the bulk electronic structure of LiFeAs. The state which has been erroneously identified as this pocket is instead a topological surface state. The TSS is only identifiable near $Z$, where the dispersive $\alpha$ bulk-band has moved above $E_F$ at $k_{||}=0$ (this evolution of the $\alpha$ band is seen most clearly in Fig. S4). As an important consideration, Bogoliubov quasiparticle interference (BQPI) measurements on LiFeAs have identified the TSS as a bulk $\alpha'$ Fermi surface \cite{Chi_2014,Sharma_2020}. This should have important consequences for the inferences drawn from phase-sensitive measurements of the LiFeAs gap parameter. While it is unclear at this stage whether the TSS is in a topologically non-trivial superconducting phase, the possibility of a time-reversal symmetry-breaking $p$-wave gap on the TSS ought to be considered in connection to the BQPI measurements \cite{Fu_2008}.

\begin{figure}
\begin{center}
\includegraphics[width=\columnwidth]{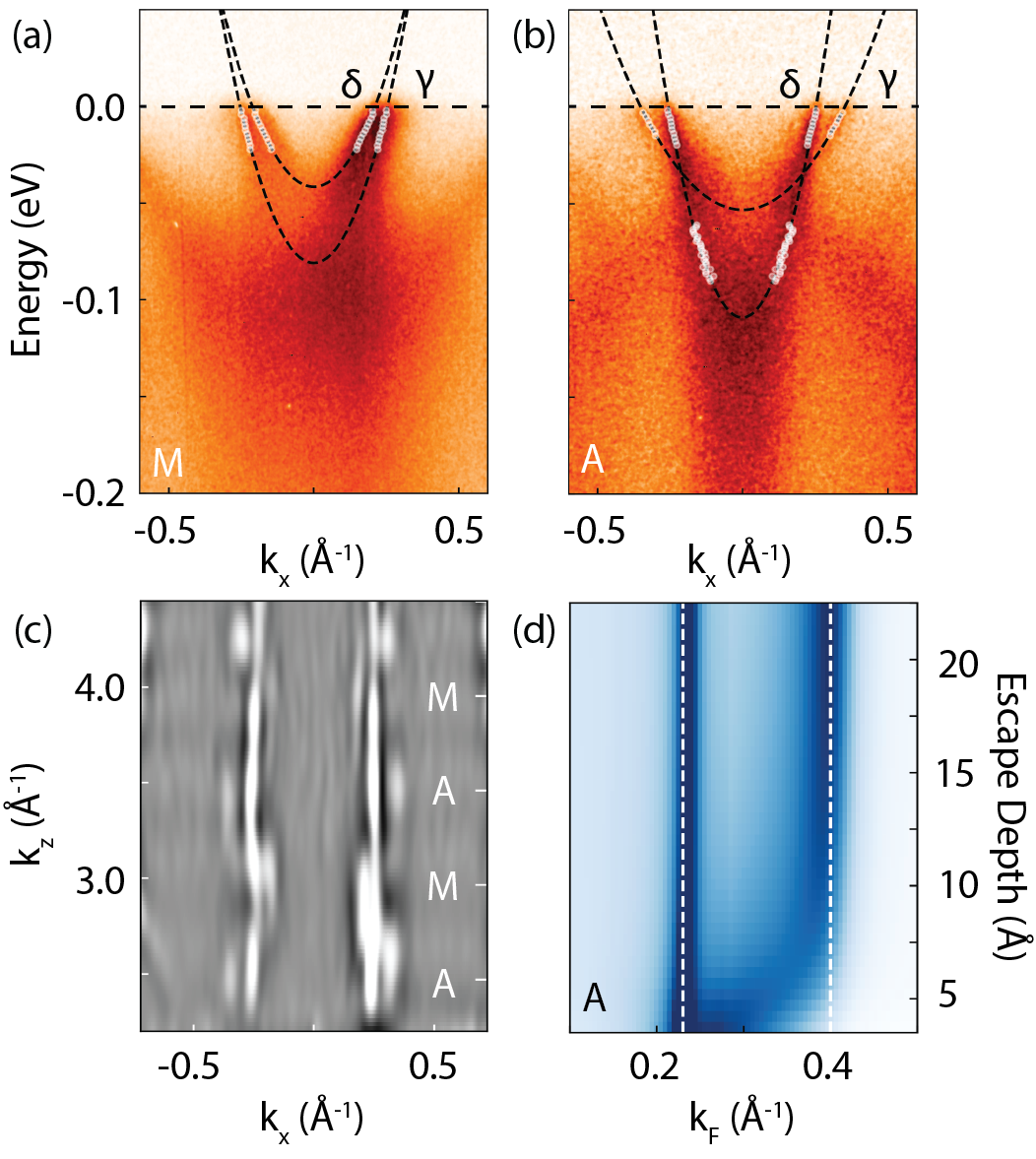}
\caption{Electron Fermi surface: low energy spectra at M ($h\nu=31$ eV (a) and A ($h\nu=44$ eV) (b) reveal the dispersive $d_{xz/yz}$ band oscillating from inside to outside the 2D $d_{xy}$ state; cursors indicate MDC peak centres and parabolic fits. Oscillation of $k_F(k_z)$ is illustrated most clearly in the curvature (as defined in Ref. \citenum{Zhang_2011}) of the Fermi surface, plot in (c) along the $k_z$ axis ($h\nu\in[18,76]$ eV). The surface sensitivity of UV-ARPES can affect the correspondence between bulk and measured $k_F(k_z)$ oscillations. This is demonstrated with simulated MDCs in (d), computed at $k_z$ = $A$ as a function of photoelectron escape depth.  
\label{fig:figMA}}
\end{center}
\end{figure}

As a final consideration, we detail the electron pockets at the Brillouin zone corner. Along the $MA$ line, we identify two electron pockets which oscillate about one another, with $k_F$ of the nearly 2D $d_{xy}$ pocket smaller at $A$ [Fig. \ref{fig:figMA}(b)] than that of the $d_{xz/yz}$ band, and larger at the $M$ point  [Fig. \ref{fig:figMA}(a)]. This interwoven dispersion is most easily recognized when plotting the curvature of the raw dataset, as in Fig. \ref{fig:figMA}(c) \cite{Zhang_2011}. We have fit this dispersion to $k_F^{xz/yz} = 0.073(3)\cdot \mathrm{cos}(k_z) + 0.282(3)\ \ang^{-1}$ and $k_F^{xy} = 0.009(1) \cdot\mathrm{cos}(k_z) + 0.251(1)\ \ang^{-1}$. The result, qualitatively compatible with predictions from DFT, differs most significantly in the amplitude of the $d_{xz/yz}$ $k_z$ dispersion. Much effort has been focused on accounting for the reduction of Fermi surface pocket areas in ARPES relative to DFT, with notable debate over the need for exotic \cite{Ferber_2012,Bhattacharyya_2020}, or more local \cite{Kim_2021} correlation effects. Although electronic correlation is undoubtedly relevant here, at least some of the apparent decrease of $k_F(k_z)$ can be understood, once again, as a consequence of surface sensitivity. This is demonstrated in Fig. \ref{fig:figMA}(d), where we plot simulated MDCs at the $A$ point of LiFeAs, calculated for various values of $\lambda$.  The dispersive $\delta$ band does not agree with its bulk $k_F$ value for $\lambda < 7$ $\ang$, below which $k_F$ rapidly decreases. This change in the apparent $k_F$ is a consequence of the reduced coordination of surface-layers; as $\lambda$ drops below the inter-layer spacing, we photoemit predominantly from these less-coordinated layers, manifest as a suppression of the $k_z$ dispersion. It is essential to note that the surface artefact of reduced $k_F$ oscillation does not influence estimation of carrier concentration via Luttinger's theorem, which has undoubtedly enabled this issue to go largely undiscussed. For such an open Fermi surface sheet, the integral over the surface volume retains the same mean value, and is then insensitive to changes to the $k_F(k_z)$ oscillation amplitudes. Although our estimation of $k_F$ is fairly consistent with quantum oscillations reports \cite{Putzke_2012,Zeng_2013}, this indicates that an ARPES-derived $k_F(M/A)$ will always reflect a lower bound estimation of bulk $k_F$ oscillations along $k_z$. Although $k_z$ uncertainty in photoemission is routinely approximated as a symmetric Lorentzian broadening over the $c$-axis of the Brillouin zone, this result is derived only by neglecting the role of the surface in the photoelectron's $z$-confinement. By introducing the surface to the calculation of photoemission intensity explicitly, as done here in the slab geometry, we may arrive at such non-trivial consequences to the $k_z$ uncertainty \cite{Day_2021}. 

One may also be concerned about the possibility of surface relaxation. If for example, the $c$-axis parameter (or rather, the interlayer spacing) increases, DFT predicts a reduction of the $k_z$ dispersion of the $d_{xz/yz}$-electron pockets. However, we confirm via surface relaxation of slab-models of LiFeAs that the surface unit cell is not appreciably modified at the vacuum interface. The only substantive change is an increase of the Fe-As bond angle, which we have verified would not reduce the apparent $k_z$ dispersion. This confirms the surface coordination as the predominant consideration in any apparent reduction of the $k_z$ dispersion of the electron Fermi surface. This analysis is detailed in the Supplementary Materials.

\section{Discussion}
The Fermi surface described by our data is summarized schematically in Fig. \ref{fig:figFS}(d). This result is qualitatively compatible with DFT predictions of the electronic structure of LiFeAs, with a few important qualifications. In addition to orbital dependent shifts of the $t_{2g}$ states to lower energy, the $k_z$ hopping amplitudes are all reduced by nearly a factor of two, with respect to the rest of the kinetic Hamiltonian. Ultimately however, the connection to DFT can be made without substantial modifications employed previously, such as the removal of the dispersive $3d_{3z^2-r^2}$ states from the low energy sector \cite{Wang_2013,Ahn_2014}. Our most controversial claim, that the $\alpha '$ band lacks a Fermi surface, is compatible with bulk probes such as the Hall and Seebeck coefficients, which indicate thermal activation of high-mobility hole carriers above approximately 100 K \cite{Tapp_2008, Rullier_2012}. 

Without an $\alpha '$ Fermi surface sheet, a description of superconductivity in LiFeAs as a Fermi surface instability mediated by intra-orbital antiferromagnetic fluctuations seems untenable. Equally important to the existence of a nesting vector is the orbital texture of the related states \cite{Graser_2009}; unlike the $\alpha '$ band, the orbital texture of the remaining $\alpha$ sheet is out of phase by $\pi/2$ with that of the electron pocket at $M/A$ and cannot support a spin-fluctuation resonance like that observed in neutron scattering experiments \cite{Qureshi_2012, Qureshi_2014, Li_2016}. Moreover, the two-dimensionality of the incommensurate peak recovered from inelastic neutron scattering cannot be related to a  3D $\alpha$ Fermi sheet, such as that identified here \cite{Qureshi_2012,Qureshi_2014}. The only likely progenitor of such a 2D peak could be the $d_{xy}$ sections of the Fermi surface, which is compatible with the conclusions of Ref. \citenum{Qureshi_2014,Li_2016}. However, the narrow magnetic bandwidth associated with fluctuations within this channel \cite{Li_2016} are unlikely to support the high $T_c$ of LiFeAs.

This situation suggests, once again, the possibility of strong coupling \cite{Umezawa_2012,Miao_2015,Lin_2021}. LiFeAs supports at least two distinct gaps, where the $2\Delta_0/k_BT_c$ ratio has been found to be 3.4 and 7.3 from STS \cite{Chi_2012}. ARPES \cite{Umezawa_2012} and optical reflectivity \cite{Dai_2016} report similar values. Although the smaller of the two (shown via ARPES \cite{Umezawa_2012} to be located on the $d_{xy}$ Fermi sheet) is compatible with the BCS prediction of 3.51, the latter cannot be reconciled with the BCS framework. In addition, the apparent strong-coupling gap is associated with the shallow $\alpha$ sheet, where $\Delta/E_F\sim 0.5$, putting at least some states in the purview of discussions around real-space pairing \cite{Chubukov_2016,Rinott_2017}. The relative success of efforts constructed on the basis of the Eliashberg theory represent a promising step in the direction of considering Fe-based superconductivity beyond weak coupling \cite{Nourafkan_2016}. Altogether, this suggests a scenario wherein the two distinct superconducting gaps in LiFeAs may descend from different physics. 

Perhaps even more exciting is the possibility of realizing topological superconductivity in the Fe superconductors, a proposal for which LiFeAs is a viable candidate \cite{Fu_2008,Zhang_2018,Zhang_2019,Konig_2019,Songtian_2020}. The charge density localized at the vacuum-interface in the TSS should be capable of supporting proximitized superconductivity, much like an engineered heterostructure, but without the complications of interfacial lattice matching. Unlike the FeSe$_{0.5-x}$Te$_{0.5+x}$ alloys which suffer from disorder challenges, the LiFeAs platform is promising due to its crystalline purity. Although substantial hybridization with the bulk $\alpha$ band limits the potential utility of the TSS in LiFeAs, strain engineering may provide the necessary orbital order to separate the TSS from the $\alpha$ band. In the Supplementary Materials, we identify that the TSS is preserved and separated from the $\alpha$-band manifold by the introduction of an on-site $\Delta_{xz/yz}$ energy scale of 10-20 meV. We note that neither identification of the TSS discussed here, nor the observation of a superconducting gap as in Ref. \citenum{Zhang_2019}, is sufficient evidence to label LiFeAs as a topological superconductor. Such a TSS can support a trivial superconducting gap under a variety of parameter values; unambiguous observation of Majorana bound states in vortex cores would be required to demonstrate a non-trivial topological state \cite{Xu_2016}. 
\section{Conclusion}
Although LiFeAs is often classified as a quasi-two dimensional material, we have presented several important consequences of its finite three-dimensionality. Coupled with the surface sensitivity inherent to ARPES, unambiguous interpretation of the electronic structure becomes a formidable challenge. This complicates estimation of $k_F$ and band-identification, as well as the evaluation of bandwidth, Fermi velocity $v_F$, and electronic self-energy \cite{Day_2021}. To this latter point, it is unsurprising that those states with greatest $k_z$ dispersion are precisely those which have been identified recently as having significantly larger orbital-specific self-energy \cite{Brouet_2016,Fink_2019}, emphasizing the need for temperature-dependent studies of scattering rates \cite{Yi_2015} to relate spectral width to correlation effects. The concepts we have explored here are not particular to LiFeAs: detailed consideration of surface sensitivity and three-dimensionality of the electronic structure are of broad importance. We have demonstrated that although most commonly reserved for explicit study of surface states, a comparison of experimental ARPES results with surface-projected electronic structure is essential to the unambiguous identification and description of spectral features. By accounting for the role of the surface in ARPES, we can more readily begin to make appropriate connections to the bulk electronic structure and properties of materials.

\section{Acknowledgements}
We are very happy to acknowledge many helpful conversations with S. R. Julian, A. Kemper, G. A. Sawatzky, M. Franz, A. Georges, M. Kim, G. Kotliar, P. Dai, M. Yi and Q. Si. This research was undertaken thanks in part to funding from the Max Planck-UBC-UTokyo Centre for Quantum Materials and the Canada First Research Excellence Fund, Quantum Materials and Future Technologies Program. This project is also funded by the Killam, Alfred P. Sloan, and Natural Sciences and Engineering Research Council of Canada's (NSERC) Steacie Memorial Fellowships (A.D.); the Alexander von Humboldt
Fellowship (A.D.); the Canada Research Chairs Program (A.D.); NSERC, Canada Foundation for Innovation (CFI); British Columbia Knowledge Development Fund (BCKDF); and the CIFAR Quantum Materials Program. Part of the research described in this work was performed at
the Canadian Light Source, a national research facility of the University of Saskatchewan, which is supported by CFI, NSERC, the National Research Council (NRC), the Canadian Institutes of Health Research (CIHR), the Government of Saskatchewan, and the University of Saskatchewan. The Flatiron Institute is a division of the Simons Foundation.

\bibliography{LiFeAs}

%
%
%%% Here is the endmatter stuff: Supplementary Info, etc.
%%% Use \item's to separate, default label is "Acknowledgements"
%
%\begin{addendum}
% \item Put acknowledgements here.
% \item[Competing Interests] The authors declare that they have no
%competing financial interests.
% \item[Correspondence] Correspondence and requests for materials
%should be addressed to A.B.C.~(email: myaddress@nowhere.edu).
%\end{addendum}
%
%%%
%%% TABLES
%%%
%%% If there are any tables, put them here.
%%

\end{document}